# ANALYSIS OF THREE NON-IDENTICAL JOSEPHSON JUNCTIONS BY THE METHOD OF LYAPUNOV EXPONENTS CHARTS

Alexander P. Kuznetsov[a], Igor R. Sataev[a], Pavel V. Kuptsov[a,b], Yuliya V. Sedova[a]

[a] Kotelnikov Institute of Radioengineering and Electronics of RAS, Saratov Branch, Zelenaya str., 38, Saratov, 410019, Russia

[b] Yuri Gagarin State Technical University of Saratov, ul. Politekhnicheskaya 77, Saratov, 410054 Russia

E-mail address: sedovayv@yandex.ru (Yu.V. Sedova)

**Abstract**

A system of three non-identical Josephson junctions connected via an RLC circuit is considered. The method of Lyapunov exponents charts is used, which makes it possible to identify the main types of dynamics of the system and to analyze the dependence of its properties on parameters. The possibility of both two and three-frequency invariant tori is demonstrated. Saddle-node bifurcations of resonant tori are studied with the use of instantaneous Lyapunov exponents. The dependence of the charts on the type of coupling in the system is discussed.

## 1. Introduction

The Josephson effect is widely used for both generating and receiving very high frequency signals. There is an extensive literature devoted to Josephson junctions (for example, [1-18]). The approaches and methods of nonlinear dynamics are commonly used: construction of phase portraits, bifurcation analysis, Kuramoto model approach, etc. Usually large arrays of identical junctions are considered. However, the non-identity of junctions leads to new and interesting effects. Two non-identical junctions with a connection via a capacitance are described in [10-11], and via a resistor - in [2,12-15]. Complex dynamics are discovered, including chaos and hyperchaos.

A very popular model is a chain of junctions connected via an RLC circuit, as shown in Fig. 1 [4-6,17]. In this case, two types of non-identity are possible - in terms of critical currents $I_n$ through the junctions and in terms of the value $r_n$ of junction resistances. An increase in the number of junctions from two to three leads to the possibility of quasiperiodic dynamics with invariant tori of dimensions both two and three and to a complex structure of the parameter space. In [18], a system of three junctions with non-identical critical currents is considered. In this paper, we will mainly consider three junctions that are not identical in terms of resistance. As the main research tool, we will use the method of construction of charts of Lyapunov exponents. Within the framework of this method, the type of dynamics of the system is determined by the signature of the spectrum of Lyapunov exponents [19-25]. The parameter plane is scanned and the types of modes are identified at each point. The method is effective in that it allows one to study all types of possible regimes and fine details of the parameter space arrangement.

One-parameter Lyapunov charts are most suitable for a more elaborated analysis of the bifurcations. An interesting type of bifurcations which is typical for the systems of coupled oscillators under study is saddle-node bifurcation of resonant tori. To reveal the peculiarities and to gain a deeper insight into the nature of this type of bifurcations we will study the behavior of so-called local instantaneous Lyapunov exponents [27] in the vicinity of the bifurcation point.

The work is structured as follows. In Sec. 2, we derive a system of equations for a chain of coupled Josephson junctions connected via RLC circuit. In Sec. 3 and 4, we discuss the structure of the parameter plane for three coupled junctions identical in critical current and non-identical in resistances. In Sec. 5, we present the results of the analysis of saddle-node bifurcation of tori using instantaneous Lyapunov exponents. In Sec. 6, we consider the case of the chain of junctions non-identical in critical currents. In Sec. 7, we discuss the effect of different types of coupling on the dynamics of the system.

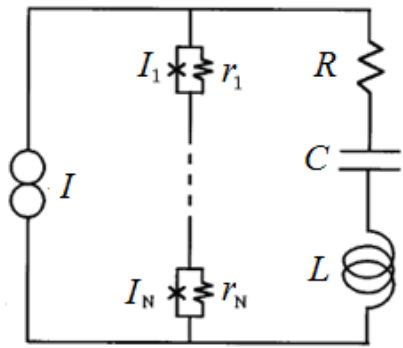

Fig. 1. Model of a chain of N Josephson junctions connected via RLC circuit [4-6].

## 2. Equations

When the wave functions for each of the superconductors forming the junction do not depend on time and are characterized by its own phase, then the equation for N junctions has the form [4-6]:

$$\frac{\hbar}{2er_n}\dot{\varphi}_n + I_n \sin\varphi_n = I - \dot{Q},$$

$$L\ddot{Q} + R\dot{Q} + Q/C = \frac{\hbar}{2e}\sum_{n=1}^{N}\dot{\varphi}_n. \quad (1)$$

Here $\hbar$ – Planck's constant, $e$ – electron charge, $I_n \sin\varphi_n$ – superconducting current, $I_n$ – critical current through the corresponding junction with the resistance $r_n$, $R$, $L$, $C$ – elements of the circuit in which the junctions are included, $I$ – external current, $\dot{Q}$ – current through parallel RLC load.

We introduce the change of variables

$$Q = \alpha Q^*, t = \beta t^*, I = I_1 I^*, \quad (2)$$

where

$$\alpha = \frac{\hbar^2 N}{4e^2 I_1 L r_1}, \beta = \frac{\hbar}{2e I_1 r_1}. \quad (3)$$

Then we obtain the following dimensionless equations

$$\frac{r_1}{r_n}\dot{\varphi}_n + \frac{I_n}{I_1}\sin\varphi_n = I - \varepsilon\dot{Q},$$

$$\ddot{Q} + \gamma\dot{Q} + \omega_0^2 Q = \frac{1}{N}\sum_{n=1}^{N}\dot{\varphi}_n. \quad (4)$$

Here

$$\varepsilon = \frac{N}{L}\frac{\hbar}{2eI_1}, \gamma = \frac{\hbar R}{2eI_1 L r_1}, \omega_0 = \frac{\hbar}{2eI_1 r_1 \sqrt{LC}}. \quad (5)$$

To shorten the notation, we omit the asterisk for new variables. We normalize the parameters and variables to the critical current $I_1$ and resistance $r_1$ of the first junction; it is clear that you can choose any resistance and current.

### 3. Dynamics of junctions that is not identical in resistance

Consider three coupled junctions identical in critical current and non-identical in resistances. Then from (4) we obtain

$$\dot{\varphi}_1 = I - \sin\varphi_1 - \varepsilon\dot{Q},$$
$$\dot{\varphi}_2 = \eta_1(I - \sin\varphi_2 - \varepsilon\dot{Q}),$$
$$\dot{\varphi}_3 = \eta_2(I - \sin\varphi_3 - \varepsilon\dot{Q}), \quad (6)$$
$$\ddot{Q} + \gamma\dot{Q} + \omega_0^2 Q = \frac{1}{3}\sum_{n=1}^{3}\dot{\varphi}_n.$$

Here $\eta_1 = \frac{r_2}{r_1}, \eta_2 = \frac{r_3}{r_1}$ – non-identity parameters for junction resistances. The case $\eta_1 = \eta_2 = 1$ responds to identical junctions.

Figure 2 shows Lyapunov exponents chart of the system (6) on the plane of non-identity parameters $\eta_1, \eta_2$ for different values of the coupling parameters ε. The values of the remaining parameters here and below $I = 1.1, \gamma = 1, \omega_0^2 = 1.2$. The parameter of the current through the junctions $I$ is chosen to be greater than unity, because otherwise, in accordance with (6), a trivial mode of stable equilibrium is observed for the phases. The following designations are used: P - periodic mode, 2T - two-frequency tori, 3T - three-frequency tori, C - chaos. The mode type was determined by the signature of the spectrum of Lyapunov exponents in accordance with the table.

At ε=0.3 for $\eta_1 > 1$ and $\eta_2 > 1$, domains of chaotic dynamics with built-in regions of two-frequency tori and periodic regimes dominate. In the case of $\eta_1 < 1$ or $\eta_2 < 1$, regions of three-frequency quasiperiodic dynamics prevail. Their domain is crossed by bands of two-frequency tori of different widths, which form a complex fractal-like structure. This structure is often referred to as Arnol'd resonant web [19-24]. Note that the widest bands are located in the vicinity of the lines $\eta_1 = 1$ and $\eta_2 = 1$, that corresponds to the identity of the first and the second or the second and the third junctions. It can also be seen from the figures that when all three junctions are identical, $\eta_1 = \eta_2 = 1$, only periodic modes are observed in the case under consideration.

With an increase in ε to a value of ε=0.5, the chaotic region expands, Fig. 2b. For $\eta_1 > 1$, $\eta_2 > 1$ small bands of two-frequency tori are destroyed. Chaos becomes possible for $\eta_1 < 1$ or $\eta_2 < 1$.

At ε=0.7, the domain of chaos expands even more and penetrates into the region where both parameters of non-identity are less than unity, Fig. 2c. In the bands of two-frequency tori in the upper right part of the map, the areas of built-in periodic modes are noticeably expanded.

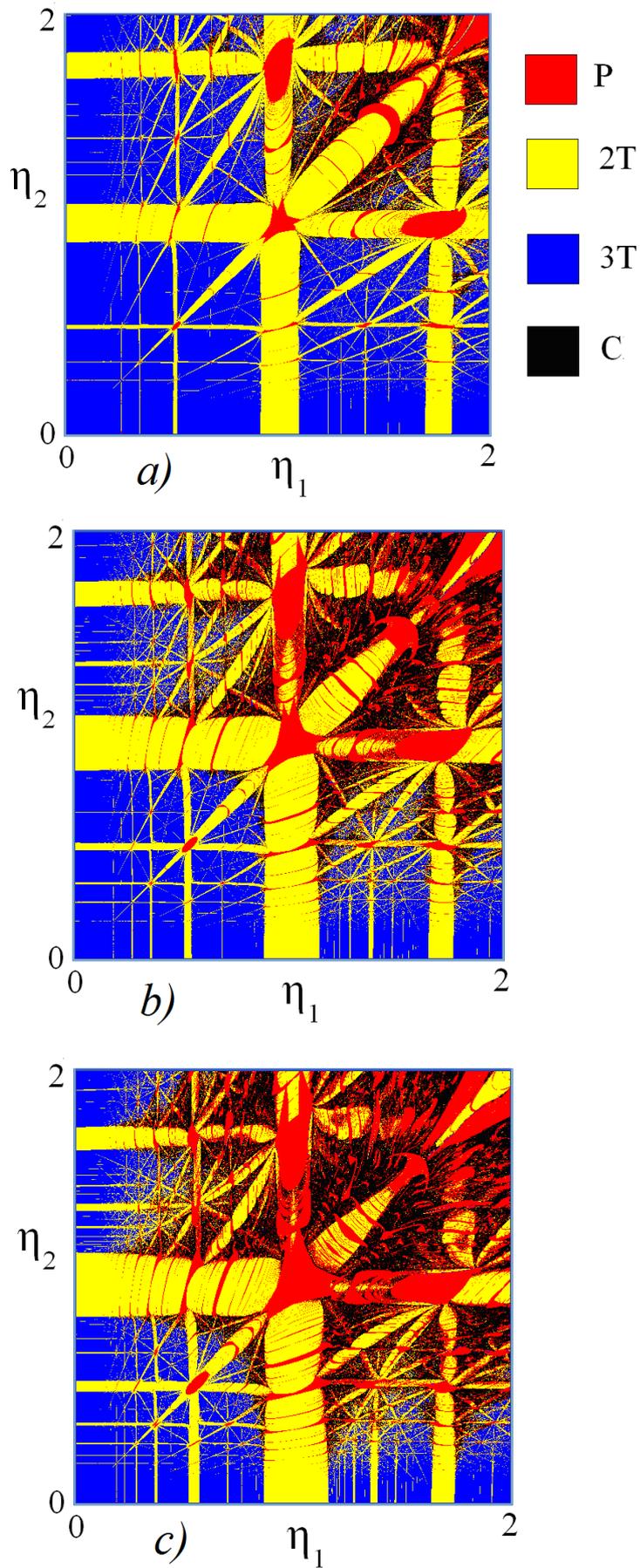

Fig. 2. Lyapunov exponents charts of system (6) on the plane of nonidentity parameters, $I=1.1$, $\gamma=1$, $\omega_0^2=1.2$; *a)* $\varepsilon=0.3$, *b)* $\varepsilon=0.5$, *c)* $\varepsilon=0.7$.

**Table 1. Correspondence between the signature of the spectrum of Lyapunov exponents, lettering and color designations**.

| Letter | Signature of the spectrum of Lyapunov exponents ($\Lambda_1$, $\Lambda_2$, $\Lambda_3$, $\Lambda_4$, $\Lambda_5$) | Color |
|---|---|---|
| P | (0, -, -, -,-) – periodic self-oscillations | magenta |
| 2T | (0, 0, -, -,-) – two-frequency quasiperiodic self-oscillations | yellow |
| 3T | (0, 0, 0, -,-) – three-frequency quasiperiodic self-oscillations | blue |
| C | (+, 0, -, -,-) – chaotic self-oscillations | black |

### 4. Arnol'd Resonant Web

Figure 3 shows a fragment of the map in Figure 2a, presenting an enlarged view of Arnol'd resonant web. One can see many bands of two-frequency tori, at the intersections of which small regions of periodic regimes are observed.

An interesting question is the type of bifurcations of invariant tori (quasiperiodic bifurcations). The answer can also be obtained using Lyapunov exponents. The graphs of the Lyapunov exponents for $\eta_1 = 1.65$ are shown in Fig. 4. The arrows show the points at which one of the Lyapunov exponents vanishes when passing from a 2T torus to a 3T torus. The form of the graphs in accordance with the classification [26] allows us to conclude that there is a saddle-node bifurcation of invariant tori, when a stable and an unstable 2-torus merge and a stable 3-torus is formed.

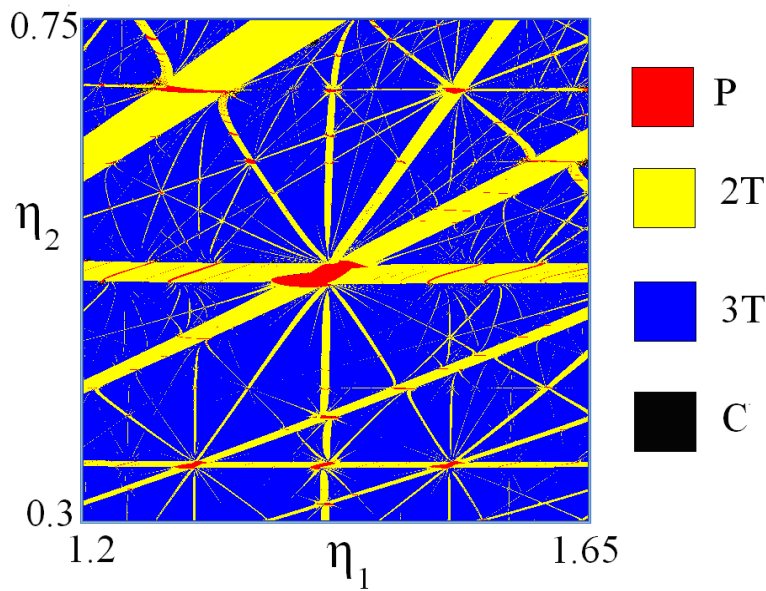

Figure 3. Arnol'd Resonant Web, ε=0.3.

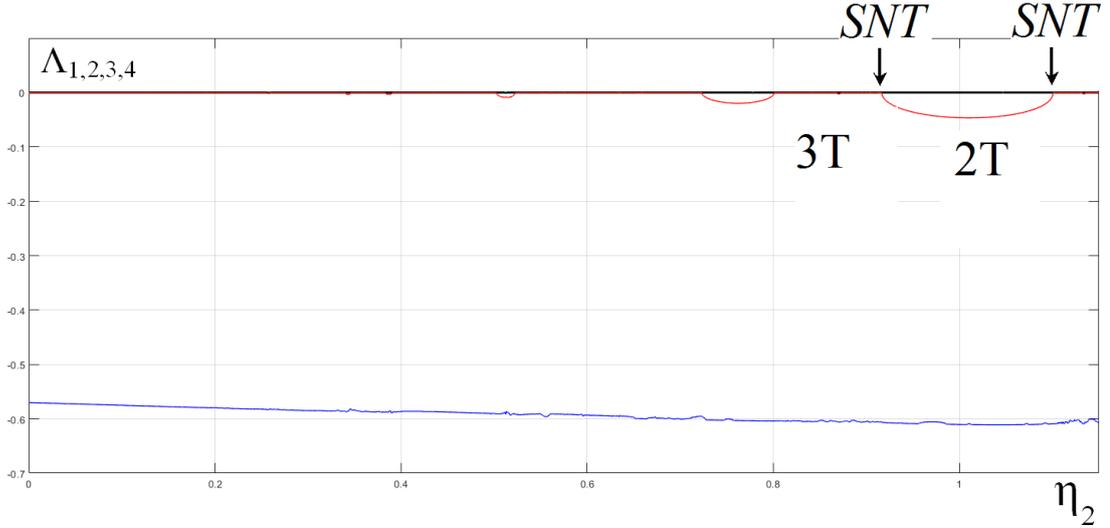

Fig. 4. Graphs of Lyapunov exponents. Arrows indicate saddle-node bifurcations of invariant tori *SNT*; ε=0.3, $\eta_1 = 1.65$.

**5. Analysis of saddle-node bifurcation of tori using instantaneous Lyapunov exponents**

Let us perform a more detailed analysis of the saddle-node bifurcation of invariant tori in the system under consideration. To do this, we apply an approach based on the calculation of the so-called local Lyapunov exponents.

When calculating the Lyapunov exponents, the accumulation and averaging of values that characterize the local stretching and compression of infinitesimal phase volumes is performed. We can also speak about local compression and stretching of volumes in tangent spaces constructed near each point of the system trajectory. We will calculate these quantities explicitly and consider their fluctuations.

There are several ways to calculate local exponents of stretching and compression. One of them, the most obvious, is to consider directly the values that are obtained during the application of the standard algorithm for calculating Lyapunov exponents. Such exponents in the literature are often referred to by the abbreviation FTLE, Finite Time Lyapunov Exponents. However, as discussed in [27], a more subtle tool, in the sense of greater sensitivity to the peculiarities of the system's behavior, is the instantaneous Lyapunov exponents calculated on the basis of covariant Lyapunov vectors.

Covariant Lyapunov vectors are calculated when moving along the trajectory of the system and show the tangent directions of its stable and unstable manifolds. In [28], theoretical and algorithmic issues of calculating these vectors are discussed. Lyapunov exponents are the average exponential growth or decline of these vectors.

In principle, they can be calculated by monitoring the behavior of covariant vectors, see [28], but this approach is quite inefficient from a practical point of view.

Moving along the trajectory, we can find local exponents of exponential growth or decrease of covariant vectors. For example, by performing discrete steps over time, as is usually the case with numerical counting, we can calculate how many times each of the covariant vectors has grown in one counting step. Then the logarithms of these quantities will be finite-time exponents calculated on covariant vectors. However, with this approach, the exponents are averaged over the time of the step. This is not very good, since the characteristic values turn out to depend on a technical parameter, a time discretization step that is not related to the physics of the system under study. Therefore, in the work [27], the so-called instantaneous Lyapunov exponents were proposed and theoretically justified. This is the limit to which the exponents averaged over a finite time tend to reach when the time interval tends to zero. We will call these indicators ICLE, Instant Covariant Lyapunov Exponents.

In various regions of the attractor, compression and stretching of phase volumes occur in different ways. The consequence of this is the fluctuations of local exponents, which can be observed by calculating them when the system moves on the attractor. During the bifurcation rearrangement of the attractor, the nature of fluctuations of local exponents should obviously change. Let the system have an invariant 2T torus and the parameter values are such that the system is located near the point of the saddle-node bifurcation. This means that in the phase space in the immediate vicinity of the stable torus there is a saddle invariant torus whose trajectories have an unstable manifold. This manifold will be located near a stable torus. Due to the continuity of the phase flows, the trajectories on the stable torus near which this unstable manifold is located will be characterized by a relatively large local stretching. As we approach the bifurcation point, this stretching will increase. In other words, when approaching the bifurcation point, the amplitude of fluctuations of local exponents should increase.

Let us turn to the system under study. Figure 5a shows the dependence of the first three Lyapunov exponents on $\eta_2$ at $\eta_1 = 1.2$. The other parameters are the same as in Fig. 2a. It can be seen that in the central region $\lambda_3 < 0$ - this corresponds to the 2T torus regime. This area is highlighted with a gray fill. Saddle-node bifurcations of invariant tori leading to the birth of a 3T torus occur at the left and right shift from the center when $\lambda_3$ turns to zero. In Fig.5b and 5c we show how the smallest and largest local ICLE exponents change, respectively. To construct these curves, for each value $\eta_2$ ICLE are calculated along sufficiently long trajectories and the minimum and maximum values are determined. The resulting dependencies are subject to strong fluctuations – even a small shift in the parameter leads to a strong change in the observed value. Therefore, in order to clearly demonstrate the behavior of the curves, we pre-processed them with a median filter.

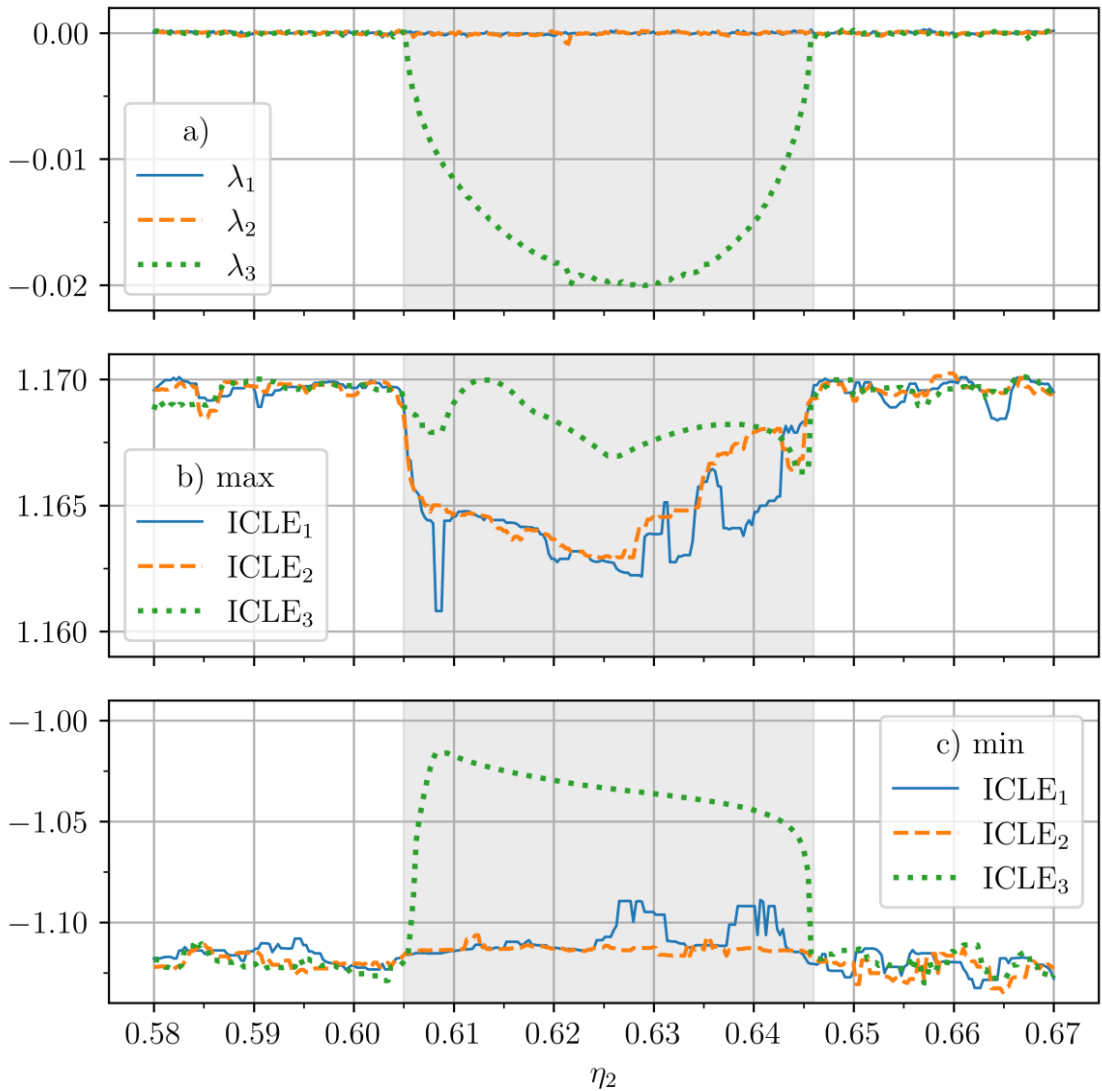

Fig.5. Lyapunov exponents, diagram a), and maximum and minimum local ICLE Lyapunov exponents, b) and c). The area highlighted in gray corresponds to $\lambda_3 < 0$.

As can be seen from Fig.5b and 5c, the amplitude of fluctuations of all three considered ICLE increases with the transition from 2T torus to 3T. For the first two exponents this is mainly due to an increase of the maximum value, see Figure 5b. Note that in accordance with the above reasonings, the increase of the amplitude of fluctuations does not occur abruptly, but more or less smoothly, as we approach the bifurcation point, that is, as the stable and saddle 2T tori approach. We also note that despite the increase of the amplitude of fluctuations, their average, that is, the first and second Lyapunov exponents, remain equal to zero, see Figure 5a, that is, they do not respond to bifurcation. The amplitude of fluctuations of the third local exponent increases due to a decrease of the minimum value. At the same time, the average, that is, the third Lyapunov exponent, increases and reaches zero.

Thus, we see that when approaching the point of the saddle-node bifurcation of invariant tori, the amplitude of fluctuations of local ICLE exponents increases. This can serve as an indicator of approaching the bifurcation point. Note that the amplitude of fluctuations varies, including the largest exponents, the average values of which remain unchanged when passing the bifurcation point.

### 6. Non-identical in critical currents junctions

Let us now discuss the influence of non-identity with respect to critical currents. For junctions identical in resistance, from (4) we obtain

$$\dot{\varphi}_1 = I - \sin\varphi_1 - \varepsilon\dot{Q},$$
$$\dot{\varphi}_2 = I - \xi_1 \sin\varphi_2 - \varepsilon\dot{Q},$$
$$\dot{\varphi}_3 = I - \xi_2 \sin\varphi_3 - \varepsilon\dot{Q}, \qquad (7)$$
$$\ddot{Q} + \gamma\dot{Q} + \omega_0^2 Q = \frac{1}{3}\sum_{n=1}^{3}\dot{\varphi}_n.$$

Here $\xi_1 = \frac{I_2}{I_1}, \xi_2 = \frac{I_3}{I_1}$ – parameters characterizing non-identity of junctions with respect to critical currents.

Figure 6 shows a map of Lyapunov exponents on the plane of nonidentity parameters $\xi_1, \xi_2$ for the parameter values used above $I=1.1$, $\gamma=1$, $\omega_0^2 = 1.2$.

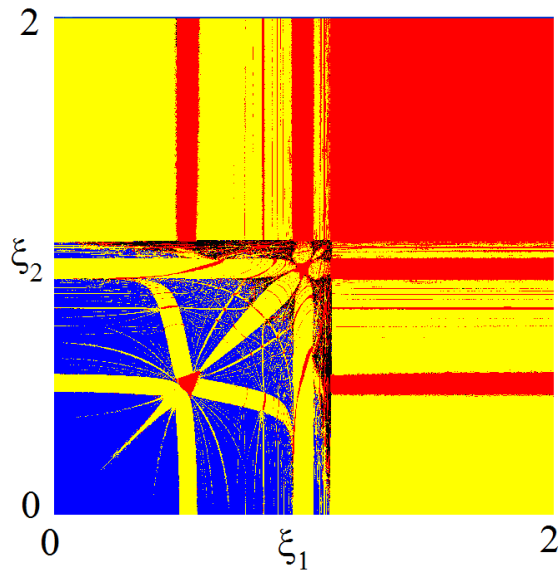

Fig. 6. Lyapunov exponents chart for three connected junctions that are not identical in critical currents (7); $\varepsilon=0.5$, $I=1.1$, $\gamma=1$, $\omega_0^2 = 1.2$.

A large region of periodic regimes is observed, the appearance of which is due to the conditions following from (7) $I < \xi_1, I < \xi_2$, when, in the absence of coupling, stable equilibrium states are observed for the second and third junctions.

You can also see that the area of three-frequency tori became smaller and is located in the lower left part of the map. Areas of chaos are practically absent. A more detailed discussion of this case can be found in [18].

Let us now give illustrations for the case when both types of non-identity – for resistances and critical currents – are present. Then from (4) the equations follow

$$\dot\varphi_1 = I - \sin\varphi_1 - \varepsilon\dot Q,$$
$$\dot\varphi_2 = \eta_1(I - \xi_1 \sin\varphi_2 - \varepsilon\dot Q),$$
$$\dot\varphi_3 = \eta_2(I - \xi_2 \sin\varphi_3 - \varepsilon\dot Q), \qquad (8)$$
$$\ddot Q + \gamma\dot Q + \omega_0^2 Q = \frac{1}{3}\sum_{n=1}^{3}\dot\varphi_n.$$

Here $\eta_1 = \dfrac{r_2}{r_1}, \eta_2 = \dfrac{r_3}{r_1}$, $\xi_1 = \dfrac{I_2}{I_1}, \xi_2 = \dfrac{I_3}{I_1}$.

Figure 7 shows the Lyapunov charts on the plane of nonidentity parameters $\eta_1, \eta_2$ for $\xi_1 = 0.3, \xi_2 = 0.2$. The last pair of parameters is chosen in accordance with Fig. 6 so that they correspond to the regime of three-frequency quasiperiodicity. It can be seen that the three-frequency quasiperiodicity is generally retained even when the resistances are not identical, with the exception of narrow bands of two-frequency tori.

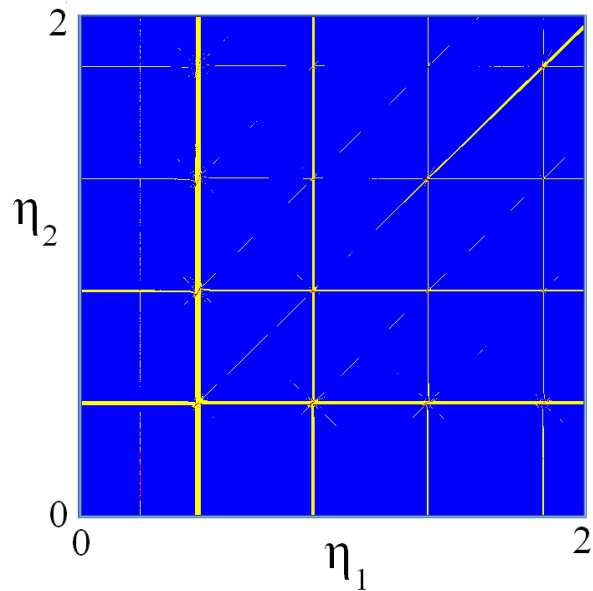

Fig. 7. Lyapunov chart for junctions that are not identical in terms of both resistance values and critical currents (8); $I=1.1$, $\gamma=1$, $\omega_0^2 = 1.2$, $\varepsilon=0.5$, $\xi_1 = 0.3, \xi_2 = 0.2$.

### 7. Effect of the coupling type

We considered the equations in the framework of the model [4-6], when the coupling between the junctions was carried out through the *RLC* circuit. Let us discuss how different types of coupling affect the dynamics of the system.

Let the external chain be an *LC*-chain, i.e. $R = 0$. Then $\gamma = 0$ according to (5). (Note that handling dimensionless equations requires a certain amount of care. For example, using the normalization [17], $\gamma = 0$ corresponds to the case when both the resistance of the external circuit and the resistances of the junctions are equal to zero.)

The Lyapunov exponents chart for the considered case on the ($\eta_1, \eta_2$) plane is shown in Fig. 8. You can see that the area of chaos is expanding. Wide areas of two-frequency tori are also observed. Arnol'd resonant web collapses.

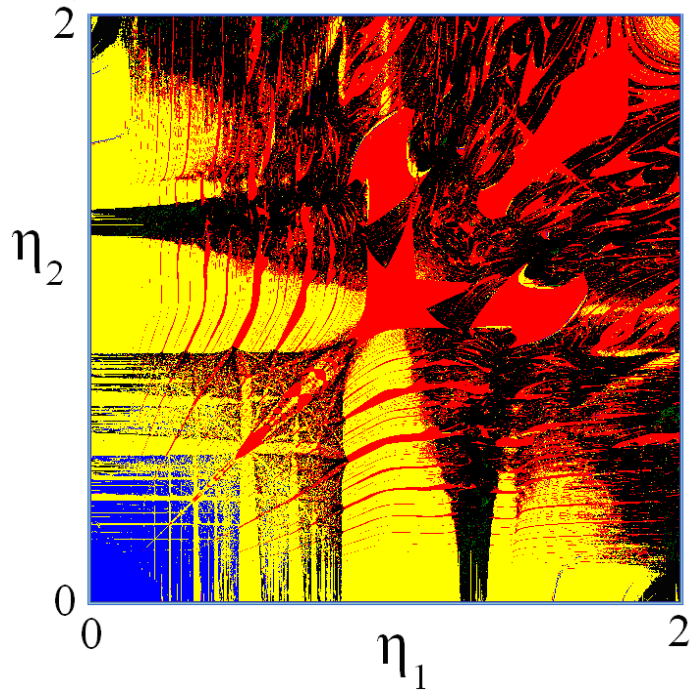

Fig. 8. Lyapunov chart for non-identical junctions (6) with coupling via an *LC* circuit; $\varepsilon=0.5$, $\gamma = 0$.

Now let the junctions are coupled through a capacitance and a resistor. In this case, normalization (3), (5) leads to a singularity at $L = 0$ and is inapplicable. We use the change of variables (2), but choose the normalization conditions

$$\alpha = \frac{\hbar N}{2eR}, \beta = \frac{\hbar}{2eI_c r_1}. \qquad (9)$$

Then we get

$$\frac{r_1}{r_n}\dot{\varphi}_n + \sin\varphi_n = I - \varepsilon\dot{Q}, \tag{10}$$

$$\dot{Q} + Q/\tau = \frac{1}{N}\sum_{n=1}^{N}\dot{\varphi}_n,$$

where $\tau = 2eI_c r_1 \dfrac{RC}{\hbar},\ \varepsilon = \dfrac{N}{R}r_1$.

For three non-identical junctions we get

$$\begin{aligned}
\dot{\varphi}_1 &= I - \sin\varphi_1 - \varepsilon\dot{Q}, \\
\dot{\varphi}_2 &= \eta_1(I - \sin\varphi_2 - \varepsilon\dot{Q}), \\
\dot{\varphi}_3 &= \eta_2(I - \sin\varphi_3 - \varepsilon\dot{Q}), \\
\dot{Q} + Q/\tau &= \frac{1}{3}\sum_{n=1}^{3}\dot{\varphi}_n.
\end{aligned} \tag{11}$$

The Lyapunov exponents chart for the system (11) at $\varepsilon=0.5$ and $\tau=0.3$ is shown in Fig. 9. The chart is qualitatively similar to the case of coupling via the *RLC* circuit, but the area of three-frequency tori now also occupies the upper right part of the map.

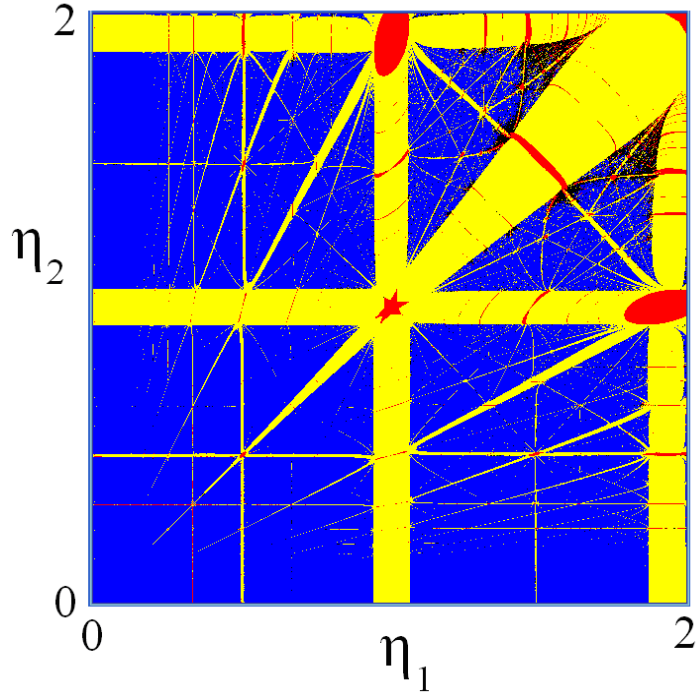

Fig. 9. Lyapunov chart for non-identical junctions (11) with coupling via an *RC* circuit; $\varepsilon=0.5$, $\tau=0.3$.

### 7. Conclusions

For the analysis of the dynamics of non-identical Josephson junctions, the method of Lyapunov exponents charts is effective. With its help, the regions of periodic regimes, regimes of two-frequency and three-frequency quasiperiodicity, chaos are revealed. As a rule, the regions of two-frequency quasiperiodicity have the form of bands of different widths immersed in the region of three-frequency quasiperiodicity, which form the structure of the Arnol'd resonant web. The boundaries of the regions of two-frequency quasiperiodicity are the lines of saddle-node bifurcations of invariant tori. As the area of chaos increases, the web can collapse. Changing the type of external circuit connecting the junctions does not fundamentally affect the dynamics of the system.

We have studied the behavior of local instantaneous Lyapunov exponents in the vicinity of the saddle-node bifurcation of invariant resonant tori. It was shown that when approaching the bifurcation point, the amplitude of fluctuations of local ICLE exponents increases. This can serve as an indicator of approaching the bifurcation point.

8. **Acknowledgements**

This work was supported by the Russian Science Foundation, project 21-12-00121, https://rscf.ru/en/project/21-12-00121/